\documentclass[12pt]{iopart}

\usepackage{iopams}

\usepackage{graphicx}
\usepackage{dcolumn}
\usepackage[colorlinks=true,linkcolor=blue ,citecolor=blue,urlcolor=blue]{hyperref}
\usepackage{multirow}
\usepackage[usenames,dvipsnames]{xcolor}
\usepackage{soul}
\usepackage{braket}
\usepackage{bm}
\usepackage{nicefrac}
\usepackage{siunitx}
\DeclareSIUnit\rydberg{Ry}
\usepackage{xcolor}
\usepackage[utf8]{inputenc}
\usepackage[T1]{fontenc}
\usepackage{upgreek}
\usepackage{tikz}

\begin{document}

\title[Interplay of magnetic states and hyperfine fields of iron dimers on MgO(001)]{Interplay of magnetic states and hyperfine fields of iron dimers on MgO(001)}

\author{Sufyan Shehada$^{1,2,3}$, Manuel dos Santos Dias$^{4,1}$, Muayad Abusaa $^{3}$ \& Samir Lounis$^{1,4}$}

\address{$^1$Peter Gr\"unberg Institut and Institute for Advanced Simulation, Forschungszentrum J\"ulich \& JARA, 52425 J\"ulich, Germany}

\address{$^2$Department of Physics, RWTH Aachen University, 52056 Aachen, Germany}

\address{$^3$Department of Physics, Arab American University, Jenin, Palestine}

\address{$^4$Faculty of Physics, University of Duisburg-Essen \& CENIDE, 47053 Duisburg, Germany}

\ead{s.shehada@fz-juelich.de \& s.lounis@fz-juelich.de}
\vspace{10pt}

\begin{abstract}
Individual nuclear spin states can have very long lifetimes and could be useful as qubits. 
Progress in this direction was achieved on MgO/Ag(001) via detection of the hyperfine interaction (HFI) of Fe, Ti and Cu adatoms using scanning tunneling microscopy (STM)~\cite{Willke2018a,yang2018electrically}. 
Previously, we systematically quantified from first-principles the HFI for the whole series of 3d transition adatoms (Sc-Cu) deposited on various ultra-thin insulators, establishing the trends of the computed HFI with respect to the filling of the magnetic s- and d-orbitals of the adatoms and on the bonding with the substrate~\cite{shehada2021trends}.
Here we explore the case of dimers by investigating the correlation between the  HFI and the magnetic state of free standing Fe dimers, single Fe adatoms and dimers deposited on a bilayer of MgO(001).
We find that the magnitude of the HFI can be controlled by switching the magnetic state of the dimers.
For short Fe-Fe distances, the  antiferromagnetic state enhances the HFI with respect to that of the ferromagnetic state.
By increasing the distance between the magnetic atoms, a transition towards the opposite behavior is observed.
Furthermore, we demonstrate the ability to substantially modify the HFI by atomic control of the location of the adatoms on the substrate.
Our results establish the limits of applicability of the usual hyperfine hamiltonian and we propose an extension based on multiple scattering processes.
\end{abstract}

%
%
\maketitle
%
%

\section{Introduction}

 Magnetic adatoms on thin insulating layers have attracted a lot of attention due to potential applications in magnetic storage~\cite{natterer2017reading,Rau2014} and quantum computation~\cite{chen2021harnessing}.
 Thin insulating layers ~\cite{Hirjibehedin2006,Hirjibehedin2007,otte2008role,Loth2010,Loth2010a,Loth2012,PhysRevLett.109.066101,PhysRevLett.112.026102,Rau2014,Baumann2015,jacobson2015quantum,Donati2016,natterer2017reading,Paul2017,Yang2017,PhysRevB.100.180405,gallardo2019large,Rejali2020} are very appealing for  qubits realizations on the basis of adatoms since their interaction with the substrate's conduction electrons is strongly reduced, which should favor coherence. 
 Magnetic adatoms on thin insulating layers such as  MgO~\cite{Rau2014,Baumann2015,Paul2017,Yang2017,Donati2016,natterer2017reading,PhysRevB.100.180405,choi2017atomic,yang2021probing}, Cu$_2$N ~\cite{Hirjibehedin2007,Hirjibehedin2006,otte2008role,Rejali2020,Loth2010,Loth2010a,Loth2012} and  h--BN~\cite{PhysRevLett.109.066101,jacobson2015quantum,gallardo2019large}  have been investigated experimentally usually using scanning tunnelling microscopy (STM) technique~\cite{Wiesendanger2009}.   
 In comparison to spin moments, individual nuclear spin states tend to have a much longer lifetime and hold in principle a greater promise as building blocks for quantum computers~\cite{mkadzik2021precision,PhysRevLett.102.257401,blinov2004quantum,PhysRevLett.117.060504,Thiele1135,vincent2012electronic}.
 Access to the hyperfine interaction coupling the nuclear spin to the electronic one~\cite{Slichter:801180} offers a wide range of applications since it provides insight into the electronic structure, magnetic state and chemical bonding of atoms, molecules and solids, as explored with nuclear magnetic resonance techniques~\cite{webb2007modern}. \\

The recent development of STM-based single-atom electron paramagnetic/spin resonance (EPR/ESR)~\cite{Balatsky2012,Baumann2015a,Choi2017a,Yang2017,Willke2018,Bae2018,Yang2019a,Willke2019a,Seifert2020,choi2017atomic,yang2021probing,seifert2021accurate}
led to several breakthroughs and enabled the possibility of detecting the weak hyperfine interaction
associated to single Ti, Fe, Cu adatoms and Ti dimer on MgO/Ag(001)~\cite{willke2019magnetic,Willke2018a,yang2018electrically}. 
We note that a deep understanding of the mechanism enabling the STM-based EPR/ESR experiment is still a puzzle under heavy investigation~\cite{Caso2014,Berggren2016,Lado2017,Reina-Galvez2019,Seifert2020,doi:10.1021/acs.jpca.9b10749,delgado2021theoretical}.
Regarding the hyperfine interactions, it is interesting that they were detected, so far, on a limited set of adatoms on a bilayer of MgO film.\\

In a previous study~\cite{shehada2021trends}, we reported on vast systematic ab initio calculations based on Density Functional Theory (DFT) of the hyperfine interactions for isolated 3$d$ magnetic transition metal adatoms (from Sc to Cu) placed on different ultrathin insulators with different thickness and bonding site, namely MgO, NaF, NaCl, h--BN and Cu$_2$N. 
We identified the adatom-substrate complexes with the largest hyperfine interactions and unveiled the main trends and exceptions.
We revealed the core mechanisms at play, such as the interplay of the local bonding geometry and the chemical nature of the thin films.
Also, we showed how the hyperfine interactions give access to information about the local electronic structure and what are the main quantities that determine their properties.\\

In this work, we take one step further by exploring the case of Fe dimers free-standing or deposited on a bilayer of MgO(001).
We note that Ti dimers were investigated experimentally~\cite{Willke2018a}.
Here we choose to study the case of Fe nanostructures because: (i) the Fe adatoms are characterized by a larger hyperfine interaction than Ti adatoms~\cite{Willke2018a,shehada2021trends}; (ii) in the experimentally measured Ti dimer, only one of the adatoms carries a nuclear spin~\cite{Willke2018a}; (iii) identification of dimer-induced multiple ESR peaks should be easier in Fe than Ti since the former carries a nuclear  spin of $\frac{1}{2}$ instead of the larger nuclear spin expected for Ti ($\frac{5}{2}$ or $\frac{7}{2}$)~\cite{bertrand2020electron}.
We investigate the impact of the magnetic alignment of the spin moments as function of their distance and bonding site on the strength of the hyperfine interaction.

\section{Methodology and computational details }

\subsection{Methodology}

The hyperfine Hamiltonian is given by~\cite{PhysRevB.35.3271,PhysRevB.47.4244}
\begin{equation}\label{eq:hypham}
    \hat{H}= \mathbf{S}\cdot\underline{\mathrm{A}}(\mathbf{R})\cdot\mathbf{I} \;,
\end{equation}
where $\underline{\mathrm{A}}$ is the hyperfine coupling tensor between the electron spin $\mathbf{S}$ and the nuclear spin $\mathbf{I}$. $\mathbf{R}$ stands for the nucleus position and the angular momenta are measured in units of $\hbar$.

The hamiltonian can be decomposed into two terms~\cite{PhysRevB.47.4244}: the Fermi contact term $\mathrm{a}$, which is isotropic and the dipolar interaction $\mathrm{b}$, being a traceless tensor: $2S\mathrm{A}_{ij}= \mathrm{a}\,\delta_{ij}+\mathrm{b}_{ij}$ ($i,j=x,y,z$).
The factor of $2S$ is used to convert between the hyperfine parameters that enter Eq.~\ref{eq:hypham} and the hyperfine field (see e.g.\ Eq.~\ref{Fermi} for the Fermi contact contribution).
Since our calculations assume the scalar-relativistic approximation and focus on the Fermi contact term, we neglect in our analysis effects induced by spin-orbit coupling and disregard the dipolar interaction.

The Fermi contact term, 
\begin{equation}\label{Fermi}
    \mathrm{a}= \frac{2P}{3}\,\rho_\mathrm{s}(\mathbf{R}) \;,
\end{equation}
finds its origin in the finite electron spin density   $\rho_\mathrm{s}(\mathbf{r})=\rho_\uparrow(\mathbf{r})-\rho_\downarrow(\mathbf{r})$ at the position of the nucleus $\mathbf{R}$. This is usually the dominant contribution. The prefactor $P = \mu_0 g_\mathrm{e} \mu_\mathrm{B} g_\mathrm{N} \mu_\mathrm{N}$, with $\mu_0$ the vacuum permeability, the electron and nuclear g-factors, $g_\mathrm{e}$ and $g_\mathrm{N}$, and the Bohr and nuclear magnetons, $\mu_\mathrm{B}$ and $\mu_\mathrm{N}$. 
The calculated hyperfine interactions are given in frequency units (e.g.\ $\mathrm{MHz}$) using Planck's constant per nuclear g-factor to facilitate comparison with future experimental data.

\subsection{Computational details}
\begin{figure}[h]
    \centering
    \includegraphics[width=\textwidth]{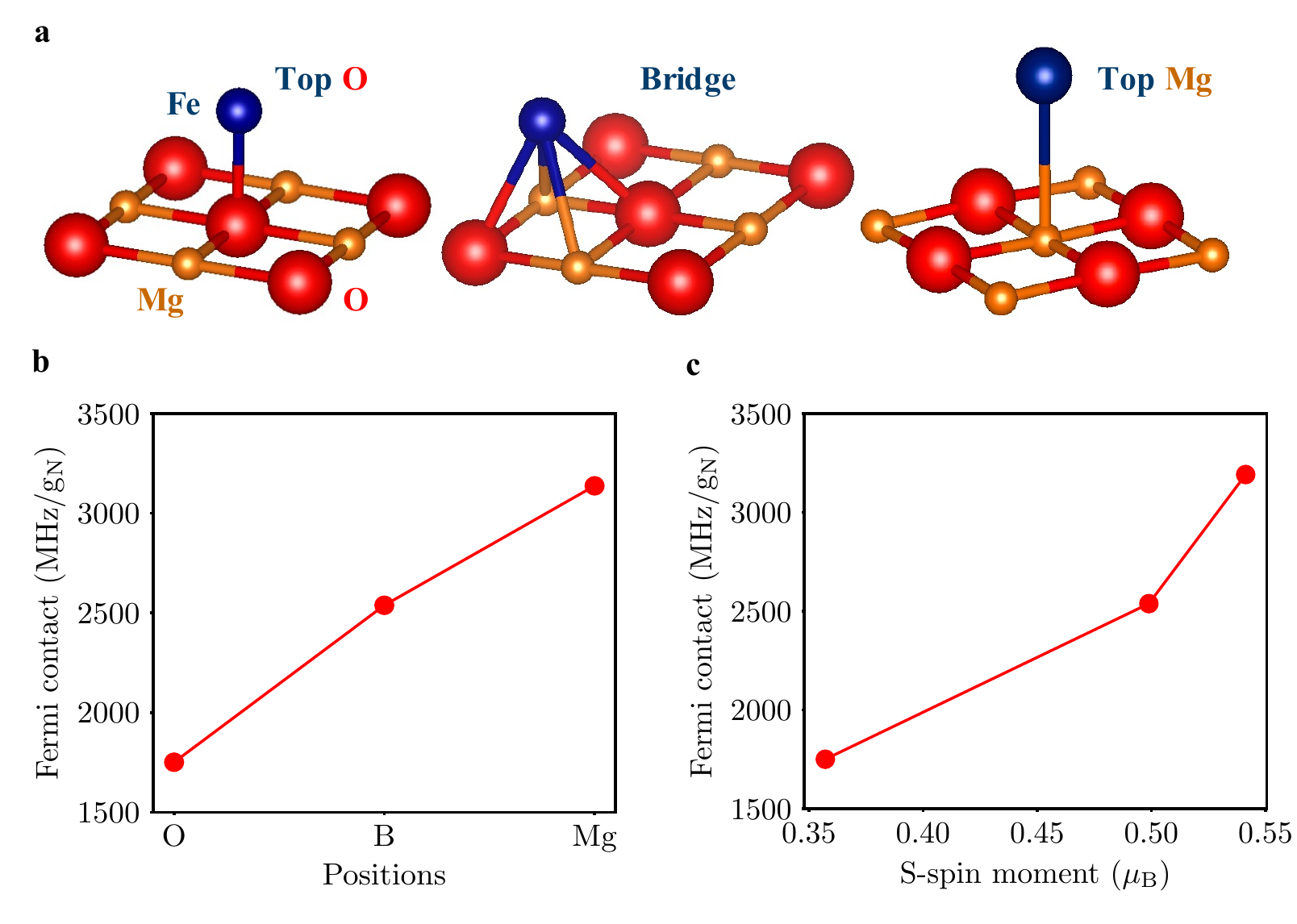}
    \caption{Fe adatoms on a bilayer of MgO.
    (a) Geometry of Fe adatoms on a bilayer of MgO.
    The second MgO layer is not shown.
    The Fe adatom is stacked on top of O, on the bridge position and on top of Mg, respectively.
    Fe is represented by a blue sphere, O by a red sphere and Mg by an orange sphere.
    (b) Fermi contact contribution to the hyperfine interaction for the considered positions and (c) Fermi contact contribution as a function of the contribution to the spin magnetic moment coming from the s electrons (S-spin moment).
    These results are recapitulated from our previous study~\cite{shehada2021trends}.}
    \label{fig:isolated_Fe}
\end{figure}
Our investigations are based on density functional theory (DFT) as implemented in the Quantum Espresso code~\cite{Giannozzi2009,Giannozzi2017}, with pseudopotentials from the PSLibrary~\cite{DalCorso2014} using the projector augmented wave (PAW) method~\cite{PhysRevB.50.17953}. 

We utilize as an exchange and correlation functional the generalized gradient approximation of Perdew, Burke and Ernzerhof (PBE)~\cite{PhysRevLett.77.3865}. 
The hyperfine interactions were evaluated with the GIPAW module of Quantum Espresso based on the theory developed by Pickard and Mauri~\cite{Pickard2001}.
For all calculations, the kinetic energy cutoff for the wavefunctions and for the charge density were set to \SI{90}{\rydberg} and \SI{720}{\rydberg}, respectively.
The Brillouin zone integrations were performed with a Gaussian smearing of width \SI{0.01}{\rydberg}.

We performed two types of simulations: The Fe dimers can be either free-standing or deposited on MgO.
For the case of free-standing dimers, we employed cubic periodic cells with a lattice constant of \SI{20}{\angstrom}, in order to minimize interactions between periodic replicas of the dimers, and $\Gamma$-point sampling of the Brillouin zone.
The noncollinear magnetism of these dimers was studied according to the theory explained in Ref.~\cite{PhysRevB.91.054420}, using ultrasoft pseudopotentials~\cite{PhysRevB.41.7892} and a bond length of \SI{2}{\angstrom}.

\begin{figure}[h]
    \centering
    \includegraphics[width=\textwidth]{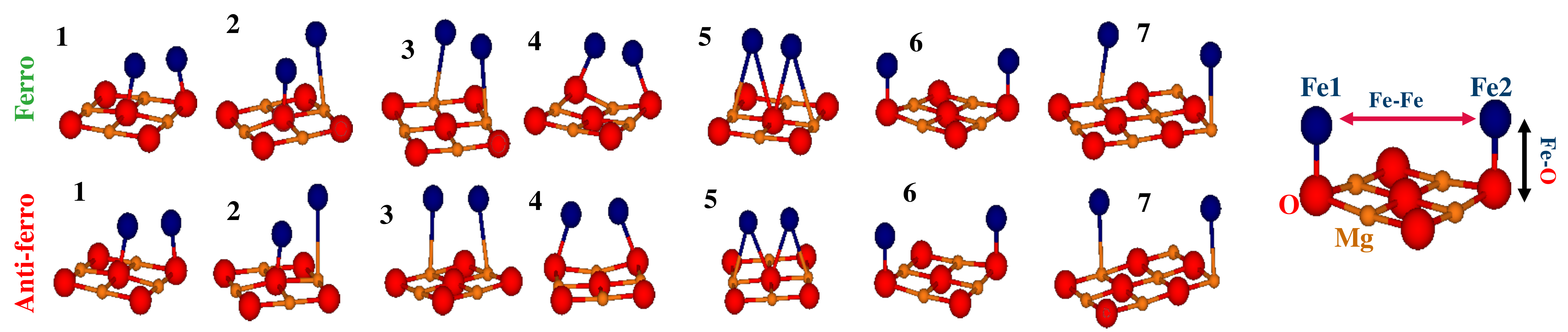}
    \caption{Atomic structures for Fe dimers on a bilayer of MgO.
    The first row represents the Fe dimers in the ferromagnetic state and the second row those in the antiferromagnetic state.
    The Fe dimer are represented by blue spheres, the O by red spheres and the Mg by orange spheres.
    The final diagram on the right indicate Fe-Fe distance represented by horizontal red double arrow line and Fe-O or Fe-Mg, whichever is nearest, represented by vertical black double arrow line}
    \label{fig:dimers_on_MgO}
\end{figure}

To accommodate the Fe adatoms and Fe dimers on MgO,
we used the theoretical lattice constants of bilayer of MgO, as obtained from our previous work \SI{4.1466}{\angstrom}~\cite{shehada2021trends}.
We then set up $3\times3$ supercells such that the Fe adatom is deposited on top of the oxygen, the bridge or the magnesium positions, as shown in (Fig.~\ref{fig:isolated_Fe}a). 
For Fe dimers on bilayer of MgO, we set the Fe adatoms on different structures as shown in (Fig.~\ref{fig:dimers_on_MgO}).
The supercells contained 73 and 74 atoms in total for the case of Fe adatoms and Fe dimers on bilayer of MgO, respectively, and a vacuum thickness equivalent to 9 layers of MgO.
We assumed a k-mesh $4\times4\times1$ in both cases (Fe adatom and Fe dimer on bilayer of MgO).
The cell dimensions were kept fixed while all atomic positions were allowed to fully relax.

\section{Results}

\subsection{Free-standing Fe dimers}

In order to set the stage for our study of the Fe dimers deposited on a MgO bilayer, we first explore the case of free-standing dimers so that the substrate and its influence is excluded.
We investigate the dependence of the Fermi contact term as function of the distance between the adatoms and of their magnetic state being ferromagnetic or antiferromagnetic. 

\begin{figure}[h]
   \centering
    \includegraphics[width=\textwidth]{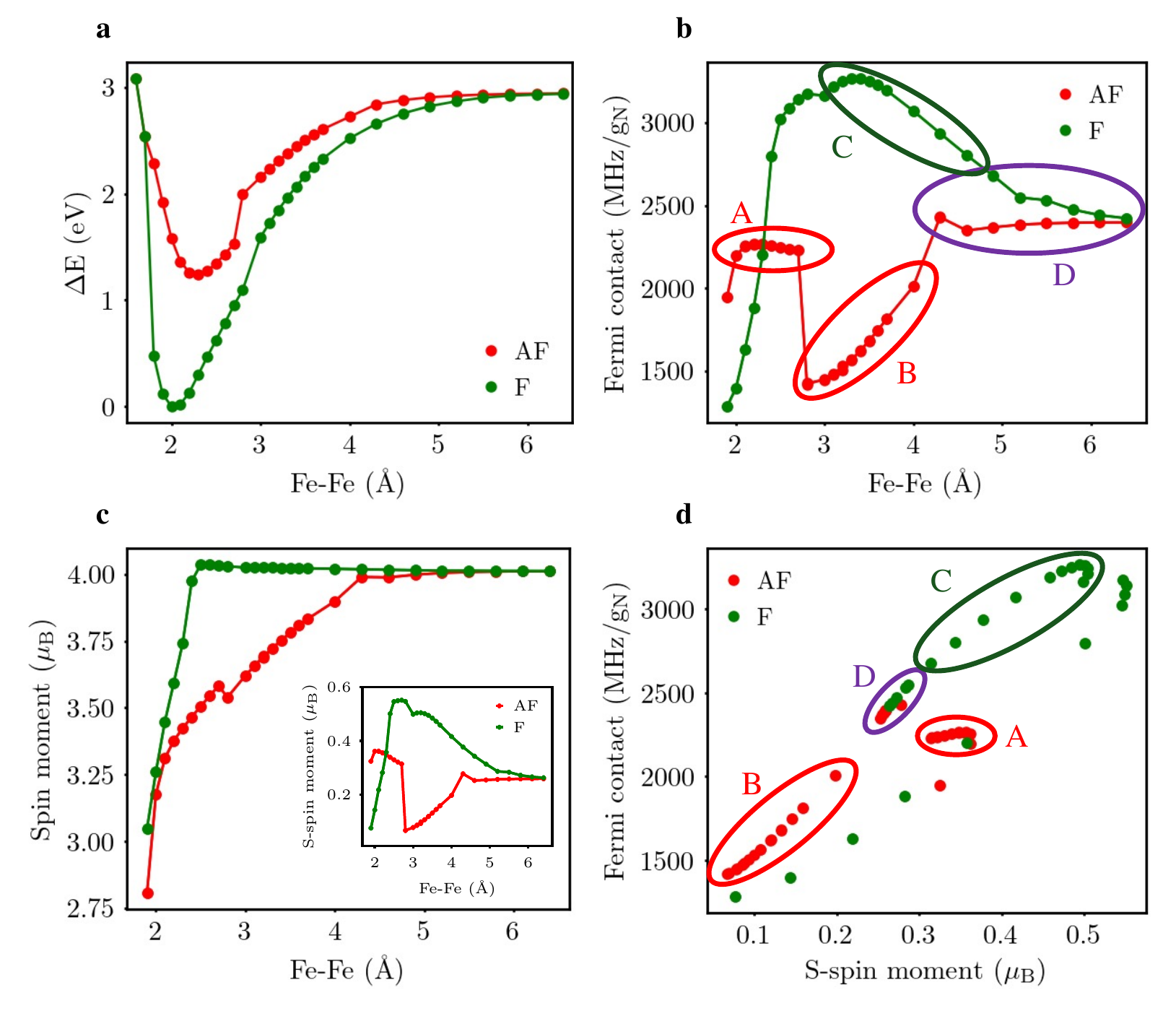}
    \caption{Basic properties of free-standing Fe dimers.
    (a) Total energy difference with respect to \SI{2}{\angstrom} of Fe-Fe distance in the ferromagnetic state, which is the most stable distance, (b) Fermi contact contribution to the hyperfine interaction, (c) atomic spin moment with the inset displaying the S-spin moment as function of atomic distances, and (d) Fermi contact contribution as a function of S-spin moment.}
    \label{fig:free}
\end{figure}

Our calculations indicate that free-standing dimers prefer to be ferromagnetic for all investigated Fe-Fe distances (up to \SI{7}{\angstrom}) as shown in Fig.~\ref{fig:free}a.
The equilibrium distance of the ferromagnetic (antiferromagnetic) dimer is \SI{2.0}{\angstrom}  (\SI{2.3}{\angstrom}) in agreement with previous theoretical work, e.g.~\cite{PhysRevB.91.054420}.

Interestingly, the dependence of the Fermi contact contribution to the hyperfine field displays a rich behavior as a function of the Fe-Fe distance, as seen in Fig.~\ref{fig:free}b.
For short distances ($\lesssim\SI{2.3}{\angstrom}$), the antiferromagnetic alignment of the spin moments induced a hyperfine field that is larger than the one obtained for a ferromagnetic state, and this behavior is reversed for larger distances.
For distances ranging from \SI{2.3}{\angstrom} till \SI{2.7}{\angstrom} defines a plateau region for the Fermi contact term of the antiferromagnetic dimers before a sharp drop leading to a minimum, 1410 $\mathrm{MHz}$, found at a distance \SI{2.8}{\angstrom}.
In strong contrast and within the same range of Fe-Fe distances, the ferromagnetic dimer reaches the maximum value of the Fermi contact term 3280 $\mathrm{MHz}$ obtained at \SI{3.3}{\angstrom}, which the distance dependence being rather smooth.
At distances larger than \SI{6}{\angstrom} the hyperfine field becomes independent of the magnetic state and approaches the value known for a single free-standing Fe atom~\cite{shehada2021trends}.

The distance-dependent behavior of the atomic spin moments for both magnetic states (Fig.~\ref{fig:free}c) is rather different from that of the Fermi contact term.
The atomic spin moment increases monotonically with the distance and for a given distance is always larger for the ferromagnetic dimer.
Its value saturates for the ferromagnetic dimer at a distance of \SI{2.5}{\angstrom} while for the antiferromagnetic one this only happens after \SI{4.3}{\angstrom}.
While the magnitude of the atomic spin moment is mostly contributed by the $d$ electrons, it is well known that the Fermi contact term is given by the spin density at the nuclear position (see Eq.~\ref{Fermi}).
The latter correlates well with the contribution of the $s$ electrons to the atomic spin moment, denoted S-spin moment, as shown in Fig.~\ref{fig:free}d and inset of Fig.~\ref{fig:free}c. 
The average trend is of proportionality between the two quantities, although the data does not fall on a single straight line and instead traces out two slightly curved lines.
Both the S-spin moment and the Fermi contact term attain their largest values for the ferromagnetic dimers, but when the S-spin moment falls below $\lesssim0.3\,\mu_\mathrm{B}$ the Fermi contact term becomes larger for the antiferromagnetic dimer than for the ferromagnetic one.

The non-trivial dependence of the hyperfine field on the magnetic state of the dimer, the distance between the atoms or the magnitude of the S-spin moment is a consequence of its sensitivity to changes in the electronic structure.
These can be identified in Fig.~\ref{fig:free}b and Fig.~\ref{fig:free}d by discontinuities or kinks, as highlighted by the labelled ovals.
In the large separation limit, the proportionality between the Fermi contact term and the S-spin moment is independent of the magnetic state, as seen in the oval labelled D in Fig.~\ref{fig:free}d, while the bonding is different for different magnetic states, which leads to a different localization of the $s$ electrons and to the differences seen for a fixed separation between the Fe atoms in Fig.~\ref{fig:free}d.

The behavior for intermediate distances evolves in opposite ways for the ferromagnetic and antiferromagnetic states, see ovals labelled B and C in Fig.~\ref{fig:free}d, while connecting to the large distance data.
This is explained by the increase (decrease) of the S-spin moment with decreasing separation for the ferromagnetic (antiferromagnetic) dimer.
In the ferromagnetic case, the atomic spin moment is constant in region C (compare with Fig.~\ref{fig:free}c), so the increase in the S-spin moment with decreasing separation is compensated by a reduction in the spin moment of the $d$ electrons.
In the antiferromagnetic case, the atomic spin moment decreases in region B (compare with Fig.~\ref{fig:free}c) by more than the decrease in the S-spin moment with decreasing separation, which also signals a reduction in the spin moment of the $d$ electrons.
The decrease in the separation between the Fe atoms leads to an increased delocalization of the $d$ electrons, which weakens the local intra-atomic exchange interaction among them and so their spin polarization.
At any distance, the $s$ electrons are much more delocalized than the $d$ electrons and so experience the combined influence of the $d$ spin moment of both Fe atoms.
With decreasing separation, the ferromagnetic alignment leads to a stronger net exchange field and so to the observed increase in the S-spin moment, while the antiferromagnetic alignment leads to a partial cancellation of the net exchange field and so the the reduction in the value of the S-spin moment.

The most striking changes happen at short separations.
In the antiferromagnetic state, the Fermi contact term jumps to a much larger value than at intermediate separations, see oval labelled A in Fig.~\ref{fig:free}b, which clearly follows from the associated jump in the magnitude of the S-spin moment, as shown in Fig.~\ref{fig:free}d, from $\lesssim0.1\,\mu_\mathrm{B}$ to $\gtrsim0.3\,\mu_\mathrm{B}$, see oval labelled A in Fig.~\ref{fig:free}d.
There is also an accompanying but smaller jump in the atomic spin moment (at \SI{2.7}{\angstrom} in Fig.~\ref{fig:free}c), but overall the polarization of the $d$ electrons is still decreasing with decreasing distance.
At the same time, the magnitude of the Fermi contact term shows a plateau behavior with respect to the value of the S-spin moment (or separation between the Fe atoms).
The magnitude of the S-spin moment is actually increasing with decreasing separation, so the plateau implies that the spin polarization at the nuclear position remains essentially constant.
In the ferromagnetic state, the S-spin moment follows the steep reduction of the atomic spin moment with decreasing, with the same behavior thus seen on the magnitude of the Fermi contact term.
However, an inspection of the data in Fig.~\ref{fig:free}d (points not encircled by a green oval) shows that the proportionality between the Fermi contact term and the S-spin moment is different at short separations than at large separations, with a smaller slope.
This is explained by a decreased spatial localization of the $s$-electrons at the nuclear position for shorter separations, as they become more concentrated in the bonding region between the Fe atoms.

The strong impact of the magnetic state provides a route for engineering the magnitude of the hyperfine interaction while at the same time raising concerns on whether the hyperfine Hamiltonian of Eq.~\ref{eq:hypham} is appropriate (for instance in combination with Heisenberg exchange interactions) to model and interpret experimental findings.
One can use multiple scattering theory as a general framework to derive how the $s$-spin density at the nuclear position of atom $i$ is expected to depend on the orientations of nearby magnetic moments, see~\ref{Appendix1}.
If one neglects spin anisotropies, we anticipate that the lowest order dependence should be proportional to the dot products of the spin moments located at sites $i$ and $j$, $\mathbf{S}_i\cdot\mathbf{S}_j$, similar to the Heisenberg exchange interaction.
Our proposed extended Hamiltonian reads: 
\begin{equation}\label{eq:hypham_new}
    \hat{H}= \sum_{i}\mathbf{S}_i\cdot\Big(\underline{\mathrm{A}}^{(0)}_i + \sum_j\underline{\mathrm{A}}^{(1)}_{ij}\,\mathbf{S}_i\cdot \mathbf{S}_j + \ldots\Big)\cdot\mathbf{I}_i 
    + \sum_{ij}J_{ij}\, \mathbf{S}_i\cdot \mathbf{S}_j \;,
\end{equation}
where $\underline{\mathrm{A}}^{(0)}_i$ is the part of the hyperfine interaction tensor which is independent of the magnetic state of the other atoms and $\underline{\mathrm{A}}^{(1)}_{ij}$ is the proposed lowest order correction, and we include the standard Heisenberg exchange interaction $J_{ij}$.

\begin{figure}[!tb]
   \centering
    \includegraphics[width=\textwidth]{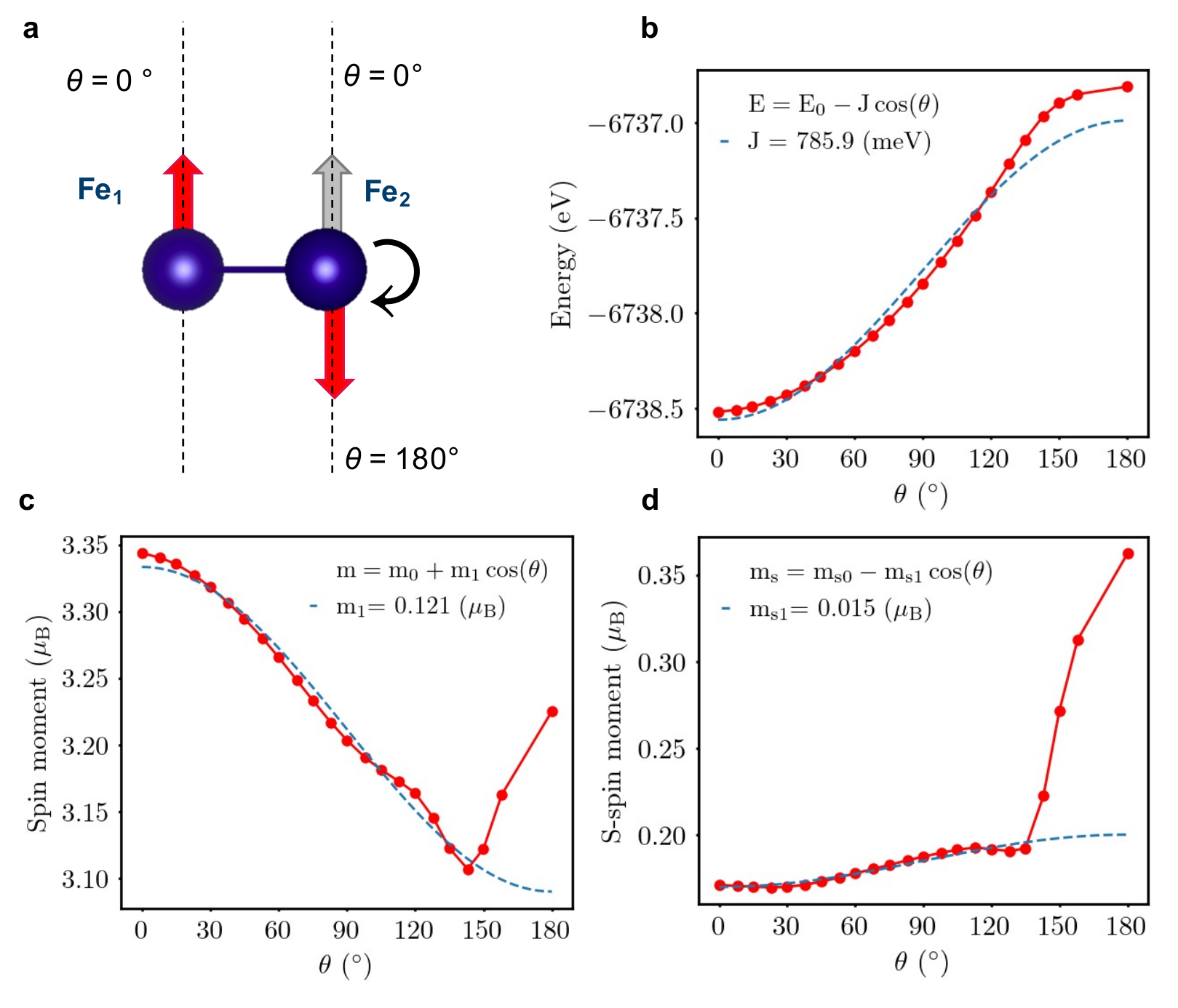}
    \caption{Noncollinear magnetism of free-standing Fe dimers.
    (a) Schematic of the considered magnetic configurations.
    (b) Total energy, (c) atomic spin magnetic moment and (d) S-spin magnetic moment as a function of the angle ($\theta$) between the directions of the spins of the Fe dimer.
    }
    \label{fig:noncollinear}
\end{figure}

We can investigate the validity of the proposed Hamiltonian by employing the S-spin moment as a proxy for the behavior of the hyperfine interaction, as already demonstrated.
We consider a free-standing Fe dimer at a fixed bond length of \SI{2.0}{\angstrom} and utilize constrained DFT to fix the spin direction of Fe$_1$ and rotate the one of Fe$_2$ (see Fig.~\ref{fig:noncollinear}a), which defines the angle $\theta$ between the two spin directions.
The total energy has an essentially $\cos\theta$ dependence (Fig.~\ref{fig:noncollinear}b), which is the expected behavior according to the Heisenberg model.

That things are not so simple is demonstrated by the angular dependence of the atomic spin moment (Fig.~\ref{fig:noncollinear}c) and S-spin moment (Fig.~\ref{fig:noncollinear}d), with their magnitudes changing in a cosine-like manner until about \SI{120}{\degree} and then evolving in a more complex way near the antiferromagnetic alignment.
This is likely due to a crossing of electronic energy levels as a function of the angle with changes in the highest occupied molecular orbital which leads to a more involved angular dependence of these key quantities.
Nevertheless, the cosine-like angular dependence holds well in two separate angular ranges, from the ferromagnetic alignment up to \SI{120}{\degree} and from there until the antiferromagnetic alignment, although with different coefficients.
This makes the proposed extended hyperfine hamiltonian Eq.~\ref{eq:hypham_new} useful for finite-temperature or non-equilibrium simulations of the ferromagnetic dimer, for instance in a pump-probe scenario.

\subsection{Recap: Hyperfine interaction of a single Fe adatom on bilayer of MgO}

Let us first recapitulate the properties of a single Fe adatom~\cite{shehada2021trends} before turning to the Fe dimers deposited on a bilayer of MgO.
Naturally, if the atoms of the dimer are sufficiently far apart one recovers the properties of the isolated adatom. 
Since they can be readily realized by atomic manipulation with STM~\cite{choi2017atomic,yang2021probing,Willke2018a}, we explored three structural scenarios: the Fe atom can sit on top of oxygen, magnesium or in the bridge position, as depicted in Fig.~\ref{fig:isolated_Fe}a.
The Fe on top of oxygen is the energetically most favorable position but it creates the weakest hyperfine field (Fig.~\ref{fig:isolated_Fe}b), followed by the bridge position before reaching a maximum when adsorbed on top of magnesium.
These are in good agreement with the trends of the S-spin magnetic moment shown in Fig.~\ref{fig:isolated_Fe}c.
The origin of the unveiled trend lies in the local bonding geometry and is closely related to the bond length between the adatom and the nearest substrate atom, indicating qualitatively the strength of hybridization of their respective electronic states.
Larger bond lengths lead to reduced hybridization, which in turn favors the localization of the spin density at the nucleus and so a larger hyperfine field.
For instance, this bond length is the shortest atop oxygen (\SI{1.9}{\angstrom}), then it increases in the bridge position (\SI{2.4}{\angstrom}) before reaching its maximum value atop magnesium (\SI{2.9}{\angstrom}).

\subsection{Hyperfine interaction of Fe dimers on a bilayer of MgO}

Here we address the last main topic of our investigation, namely dimers placed on MgO bilayer considering different location of the Fe atoms and inter-adatom distances and assuming both the ferromagnetic and antiferromagnetic states.
After structural relaxation, we classified the results into seven structures which strongly depend on the magnetic alignment of the spins, as shown in Fig.~\ref{fig:dimers_on_MgO}.

\begin{figure}[tb]
    \centering
    \includegraphics[width=\textwidth]{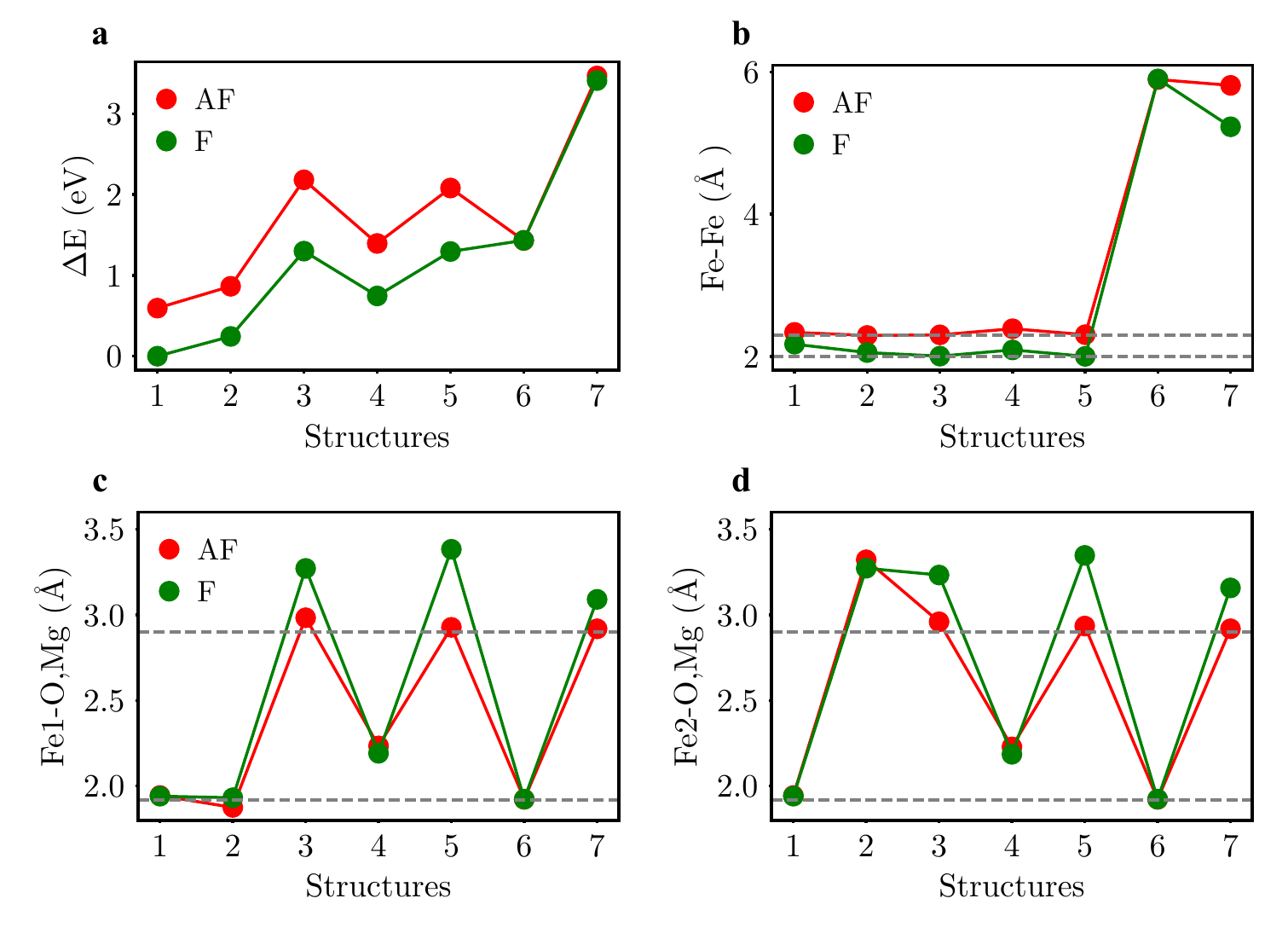}
    \caption{Relaxed structural properties of Fe dimers on the MgO bilayer.
    (a) Total energy difference with respect to structure 1 in the ferromagnetic state, which is the most stable one.
    (b) Fe-Fe distance, (c) distance between Fe1 and either O or Mg, whichever is nearest, and (d) likewise for Fe2.}
    \label{fig:Fe_geometrical_properties}
\end{figure}

As shown in Fig.~\ref{fig:Fe_geometrical_properties}a and similarly to the free-standing case, the Fe dimers prefer to be in a ferromagnetic state, with structure 1 (see Fig.~\ref{fig:dimers_on_MgO}) being the most stable.
We find that switching the atomic spin alignment from parallel to antiparallel increases the Fe-Fe distance, as illustrated in Fig.~\ref{fig:Fe_geometrical_properties}b, the exception being structure 6.
In structures 1 to 5 the Fe adatoms are close to each other with a separation only slightly larger than that found for the free-standing dimers in the respective magnetic states, while in structures 6 and 7 they are much farther apart.
The distances between each Fe atom and the nearest substrate atom are given in Fig.~\ref{fig:Fe_geometrical_properties}c and Fig.~\ref{fig:Fe_geometrical_properties}d.
Short distances indicate that oxygen is underneath while long distances signal magnesium, with the values close to but in most cases substantially larger that those found for an isolated Fe adatom.

\begin{figure}[tb]
    \centering
    \includegraphics[width=\textwidth]{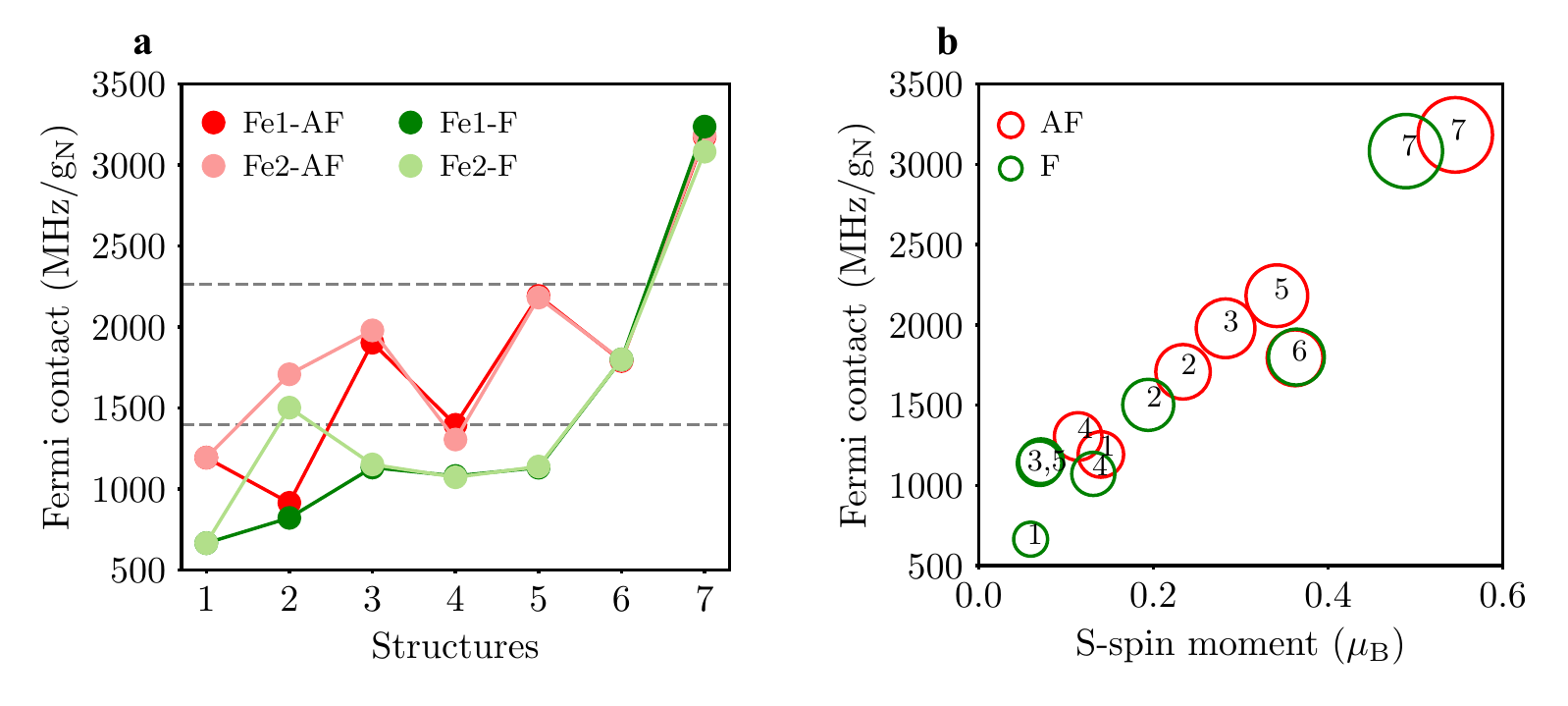}
    \caption{Hyperfine fields of Fe dimers on the MgO bilayer.
    (a) Fermi contact contribution to the hyperfine interaction for the different structures and magnetic states.
    The horizontal dashed lines indicate the values found for the free-standing dimer at their equilibrium bond length in the ferromagnetic (lower line) and antiferromagnetic (upper line) states.
    (b) Fermi contact contribution as a function of the S-Spin magnetic moment.
    The number inside the circle identifies the Fe dimer structures.}
    \label{fig:result_Fe_on_MgO}
\end{figure}

The variation of the hyperfine interaction across the different dimer structures in both considered magnetic states is plotted in Fig.~\ref{fig:result_Fe_on_MgO}a.
Given the previous discussion, it makes sense to compare structures 1 to 5 to the results for the free-standing dimers and structures 6 and 7 to the results for the isolated adatoms on MgO.
Starting with the latter two structures, we do find that the obtained values of the Fermi contact term are very similar to the adatom values and have little dependence on the magnetic state of the dimer, confirming the weak coupling between the Fe atoms (see also Fig.~\ref{fig:Fe_geometrical_properties}a).

In contrast, structures 1 to 5 show a strong coupling between the two Fe atoms combined with a strong influence of the MgO bilayer on the hyperfine interactions.
Overall the values of the Fermi contact term are substantially reduced from those obtained for the free-standing dimers in the respective magnetic states and at their equilibrium bond lengths.
This is a consequence of the dimer bonding with the MgO bilayer, which leads to an increased delocalization of the $s$ electrons and so to a reduction of the value of the Fermi contact term.
In structures 1, 3, and 5, changing the dimer from ferromagnetic to antiferromagnetic leads to an almost doubling in the magnitude of the Fermi contact term, while for structures 2 and 4 the change in the magnetic state has a much weaker impact on the hyperfine interaction.
This cannot be simply rationalized in terms of strong vs.\ weak magnetic coupling between the Fe atoms in the dimer, as Fig.~\ref{fig:Fe_geometrical_properties}a shows that for all these structures there is a large total energy difference between the two magnetic states.
Instead, it reveals in which structures the $s$ electron orbitals change strongly or weakly in response to a change in the magnetic state of the dimer.

Fig.~\ref{fig:result_Fe_on_MgO}b shows that the proportionality between the Fermi contact values and the S-spin magnetic moment is fairly independent of the structure and magnetic state of the Fe dimer.
In contrast to the results for the free-standing dimers, on MgO the data follows a single linear relation between the two quantities, which is even more apparent if structures 6 and 7 are ignored (as they are closer to the isolated adatom behavior).
This can be interpreted as a stabilization of the electronic structure against magnetic or structural changes, in particular of the $s$-states, due to the bonding between the Fe atoms and the substrate.
We thus see that the S-spin magnetic moment is an excellent proxy for the hyperfine field of the Fe dimers on MgO, and so our findings concerning the noncollinear magnetic states of the free-standing dimers should also apply in this situation.

\section{Conclusion}

We presented the results of ab initio calculations on the main contribution to the hyperfine interaction, namely the Fermi contact, of free standing Fe dimers, isolated Fe adatoms and Fe dimers deposited on a bilayer of MgO. We explored in particular their structural, electronic and magnetic properties and scrutinized the impact of the inter-adatom distance as well as the magnetic state on the hyperfine interaction, whose physics is mainly dictated by the polarization of the $s$ electrons at the nucleus position. 

We revealed the non-trivial influence of the magnetic alignment of the moments, being parallel or antiparallel, on the magnitude of the Fermi contact term. At short inter-adatom distances, the antiferromagnetic state enhances the hyperfine interaction with respect to that of the ferromagnetic state while the opposite behavior is found for large distances. This opens on the one hand the possibility of controlling the magnitude of the Fermi contact term by switching the magnetic state of the nanostructure. This could potentially be achieved via for example: (i) atomic manipulation utilizing atomic decoration, by engineering the environment of the adatoms, (ii) inelastic scanning tunneling spectroscopy or (iii) an external magnetic field. On the other hand, our findings question the use of the hyperfine Hamiltonian, eq.~\ref{eq:hypham}, usually amended with conventional Heisenberg exchange interaction to describe the interplay of magnetic coupling and hyperfine interaction, which is not anymore a constant, in multi-atomic structures. To address the latter aspect, we proposed an extended hyperfine-Heinseberg Hamiltonian, eq.~\ref{eq:hypham_new}, where the hyperfine interaction is proportional to the dot product of the spin moments, as motivated from multiple-scattering theory. Our calculations confirm that the angular dependence is reasonable for a wide range of angles around the value obtained for the ferromagnetic state.

Moreover, we evidenced the ability to substantially modify the Fermi contact term by atomic control through the control of the location of the adatoms on the substrate. The nature of the nearest neighboring surface atoms impact on the adatom-substrate distance, which affects the localization of the spin-polarized $s$ electrons and therefore the hyperfine interaction.


\section{Acknowledgements}
This work was supported by the Federal Ministry of Education and Research of Germany in the framework
of the Palestinian-German Science Bridge (BMBF grant number 01DH16027).  
We acknowledge the computing time granted by the JARA-HPC Vergabegremium and VSR commission on the supercomputer JURECA at Forschungszentrum Jülich~\cite{jureca} and RWTH Aachen University under project jara0189.

\appendix
\section{Induced magnetization from multiple-scattering expansion \label{Appendix1}}
We invoke a multiple scattering formalism based on Green functions describing the propagation of the electron states from atom $i$ to atom $j$.
The theory is similar to the one utilized to explain the spin-mixing magnetoresistance in Refs.~\cite{Crum2015,Fernandes2020,Bouhassoune2021,Fernandes2022} or higher-order magnetic interactions in Refs.~\cite{Brinker2019,Grytsiuk2020,Lounis2020,Dias2022}.
The spin density of atom $i$ is obtained from the Green function as
\begin{equation}\label{eq:spinmomgf}
  \mathbf{m}^\mathrm{spin}_i(\mathbf{r}) =  -\frac{1}{\pi}\,\mathrm{Im} \!\int\!\mathrm{d}E\;f(E) \Tr_\mathrm{spin} \boldsymbol{\sigma}\,G_{ii}(\mathbf{r},\mathbf{r};E) \;.
\end{equation}
Here $\mathbf{r}$ is the electronic position measured with respect to the nuclear position of atom $i$, $f(E)$ is the Fermi-Dirac distribution, $\boldsymbol{\sigma} = \left(\sigma_x,\sigma_y,\sigma_z\right)$ is the vector of Pauli matrices, and $G_{ii}(\mathbf{r},\mathbf{r};E)$ is the position-diagonal part of the electronic Green function at energy $E$.
The total spin moment is obtained by integrating the spin density in a suitably-defined region around atom $i$,
\begin{equation}
    \mathbf{m}_i = \int_{\Omega_i}\hspace{-0.5em}\mathrm{d}\mathbf{r}\;\mathbf{m}^\mathrm{spin}_i(\mathbf{r}) \;,\quad
    \mathbf{S}_i = \frac{\mathbf{m}_i}{|\mathbf{m}_i|} \;,
\end{equation}
from which we can also define its orientation $\mathbf{S}_i$.

The Fermi contact and the dipolar contributions to the hyperfine interaction are defined in terms of the spin density in the vicinity of the nucleus, and we now seek to establish how this quantity is modified by changes in the magnetic state of nearby atoms.
The Kohn-Sham hamiltonian can be written as
\begin{equation}
    \mathcal{H} = \mathcal{H}^0 + \sum_i B_i^\mathrm{xc}(\mathbf{r})\,\mathbf{S}_i \cdot \boldsymbol{\sigma}
    = \mathcal{H}^0 + \mathcal{H}^\mathrm{mag} \;,
\end{equation}
where $\mathcal{H}^0$ collects contributions which do not depend on the spin orientations, and the exchange-correlation magnetic field $B_i^\mathrm{xc}(\mathbf{r})$ for atom $i$ is assumed to point along the orientation of the respective total spin moment $\mathbf{S}_i$.
Here the spin-orbit interaction is neglected for simplicity.

As shown in e.g.~Ref.~\cite{Brinker2019}, we can make use of the Dyson equation to also separate the Green function as
\begin{equation}
    G = G^0 + G^0 \mathcal{H}^\mathrm{mag} G^0 + G^0 \mathcal{H}^\mathrm{mag} G^0 \mathcal{H}^\mathrm{mag} G^0 + \ldots
\end{equation}
with the reference Green function $G^0 = \left(E - \mathcal{H}^0\right)^{-1}$.
Keeping only the first magnetic term and using the definition for the spin density, Eq.~\ref{eq:spinmomgf}, we find
\begin{equation}
  \mathbf{m}^\mathrm{spin}_i(\mathbf{r}) =  -\frac{1}{\pi}\,\mathrm{Im} \!\int\!\mathrm{d}E\;f(E)
  \sum_j \int\!\mathrm{d}\mathbf{r}'\; G_{ij}^0(\mathbf{r},\mathbf{r}';E)\,B_j^\mathrm{xc}(\mathbf{r}')\,\mathbf{S}_j\,G_{ji}^0(\mathbf{r}',\mathbf{r};E) + \ldots
\end{equation}

Projecting on the orientation of the spin moment of atom $i$ and integrating out the spatial dependence in combination with the projector $\mathcal{P}_i(\mathbf{r})$ defining the hyperfine interaction (this could be to extract the Fermi contact or the dipolar contributions) we find
\begin{equation}
    \int\!\mathrm{d}\mathbf{r}\;\mathcal{P}_i(\mathbf{r})\,\mathbf{m}^\mathrm{spin}_i(\mathbf{r})\cdot\mathbf{S}_i = A_i + \sum_{j \neq i} A_{ij}\,\mathbf{S}_i\cdot\mathbf{S}_j \;.
\end{equation}
Although this result was obtained using only the lowest-order contribution, taking higher-order terms into account will only modify the definitions of the coefficients but not the dependence on the relative orientations of the spin moments.

\section*{References}
\bibliographystyle{iopart-num}
\bibliography{reference}


\end{document}